\documentclass[article]{elsarticle}

\usepackage{lineno,hyperref}
\usepackage{multirow}
\usepackage{array}
\usepackage{subfigure}

\usepackage{soul}
\usepackage{color}
\soulregister{\cite}7
\soulregister{\ref}7
\sethlcolor{yellow}

\begin{document}

\begin{frontmatter}


\title{Modeling the Charge Collection Efficiency in 
the Li-diffused Inactive Layer of P-type High Purity
Germanium Detector}

\author[THU]{W.H. Dai}
\author[THU]{H. Ma\corref{mycorrespondingauthor}}
\cortext[mycorrespondingauthor]{Corresponding author: mahao@tsinghua.edu.cn}
\author[THU]{Q. Yue}
\author[THU]{L.T. Yang}
\author[THU]{Z. Zeng}
\author[THU,BNU]{J.P. Cheng}
\author[THU]{J.L Li}

\address[THU]{Key Laboratory of Particle and Radiation Imaging 
(Ministry of Education) and Department of Engineering Physics, 
Tsinghua University, Beijing 100084}
\address[BNU]{College of Nuclear Science and Technology, 
Beijing Normal University, Beijing 100875, China}

\begin{abstract}
A model of the Li-diffused inactive layer in P-type high purity
germanium detectors is built to describe 
the transportation of charge carriers and 
calculate the charge collection efficiency therein. 
The model is applied to calculate
charge collection efficiency of 
a P-type point-contact germanium detector
used in rare event physics experiments
and validated in another 
P-type semi-planar germanium detector.
The calculated charge collection efficiency curves
are well consistent with measurements for both detectors.
Effects of the Li doping processes on the charge collection efficiency 
are discussed based on the model. 
This model can be easily extended to other P-type germanium
detectors, for instance, the P-type broad-energy Ge detector, 
and the P-type inverted-coaxial point-contact detector.
\end{abstract}


\end{frontmatter}


\section{Introduction}
With excellent energy resolution,
high detection efficiency, 
and extremely low intrinsic background, 
high purity germanium detectors (HPGe) have been  
widely used in radiation detection, 
nuclear physics, particle physics, and astrophysics \cite{bib:1,bib:2,bib:3}. 
The P-type high purity germanium detector
with a Li-diffused layer on the surface 
has been used to search for rare processes, for instance
dark matter detection \cite{bib:4}-\cite{bib:8}, 
and the neutrinoless double-beta decay \cite{bib:9,bib:10}.

Fabricated by lithium diffusion technology, 
the surface N+ layer of the HPGe detector 
has been recognized to have incomplete 
charge collection and therefore is referred to 
as an inactive or dead layer \cite{bib:11,bib:12,bib:13}. 
Measurements of the signals originating in 
the inactive layer suggest a two-layer structure: 
a total dead layer near the detector surface 
where no charge is collected, 
and a transition layer where 
charge collections are incomplete, 
induced signals are with characteristic time features \cite{bib:13,bib:14}.

For the P-type HPGe detector used in rare events searching experiments, 
the inactive layer serves as a passive shield against 
the surface contaminations of the detector but 
reduces the active volume and produces potential backgrounds from 
the incomplete charge collection signals \cite{bib:13,bib:15}. 
Therefore, knowledge of the thickness and structure of 
the inactive layer is important for correcting the 
detector’s active volume and understanding the backgrounds.

Measurement of the inactive layer thickness has been discussed
in literature \cite{bib:11,bib:12,bib:16}. 
The diffusion and recombination of charge carriers were 
considered in \cite{bib:15,bib:20} to model the 
charge collection efficiency (CCE).
In our previous works, the inactive layer of a 1-kg 
P-type point-contact Ge (PPCGe)
detector (CDEX-1B) has been studied 
\cite{bib:4,bib:16}.
A full charge collection depth (FCCD) of 
850$\pm$120 $\mu$m was measured using 
$^{133}$Ba calibration source \cite{bib:16}.
An empirical CCE function 
of the lateral inactive layer has been computed 
using calibration data and Monte Carlo simulation,
details of the construction of the empirical CCE 
function can be found in \cite{bib:16}.

In this work, a 1-D model based on first principle 
is built to calculate the CCE
in the inactive layer of the P-type HPGe detector.
Our model includes a charge carrier mobility model,
a simplified recombination model,
and calculations of the electric fields 
in the inactive layer to 
model the diffusion, drift, and recombination
of charge carriers and calculate the CCE in
the inactive layer.
Our 1-D model dose not consider the self repulsion of 
charge carriers.
The model is applied to calculate the CCE in
the inactive layer of 
the CDEX-1B PPCGe detector
and validated in a P-type semi-planar Ge detector (PSPGe).
The effects of detector fabrication
process on the inactive layer are also discussed 
based on the model.

\section{The inactive layer model of a P-type HPGe detector}
\begin{figure}[!htb]
\includegraphics
[width=1.0\hsize]
{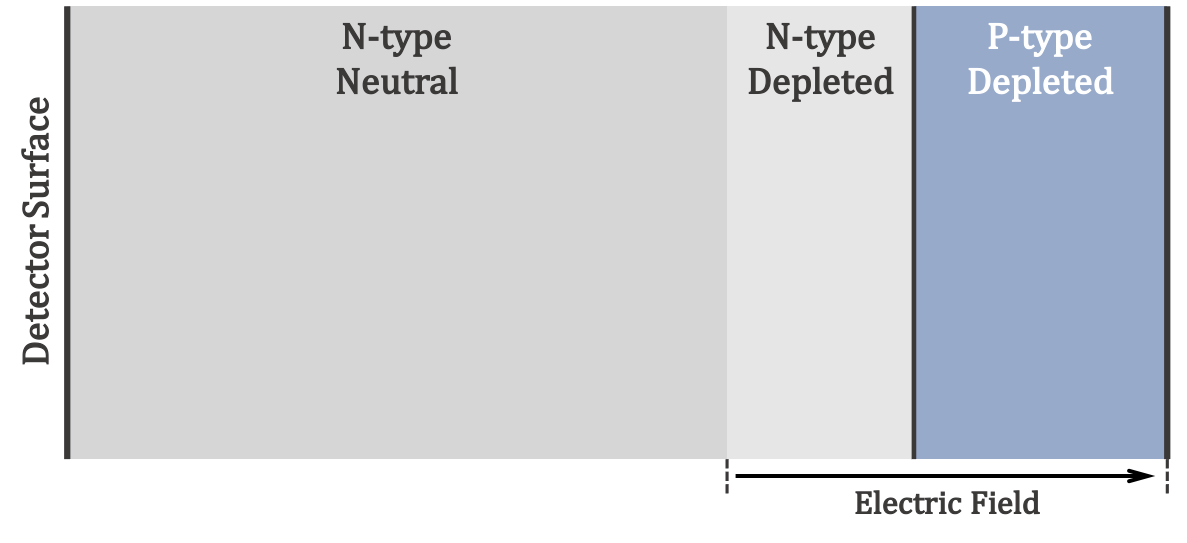}
\caption{\label{fig:DLDiag} Schematic diagram of the Li-diffused N+ layer}
\end{figure}

The N+ surface of a P-type HPGe
detector can be formed by doping donor 
impurities (Li) on the surface. 
The pn-junction is generated near the detector surface 
when a reverse bias voltage is applied. 
Full depleted regions are formed both on the P-type and N-type 
regions and leave a non-depleted neutral N-type region on the 
surface, as shown in Fig.\ref{fig:DLDiag}.

Charge carriers, holes, and electrons are created 
when energy depositions happen in the germanium crystal. 
Signals are generated by the charge carriers as 
they move in the weighted potential field of the p+ readout 
electrode \cite{bib:17,bib:18}. 
Due to the small contact size of the PPCGe detector
and the P-type semi-planar Ge detector, 
majorities of the output signals are formed by holes 
when they drift near the p+ electrode \cite{bib:19}. 
Therefore, for charge carriers generated in the 
surface N-type region,
only holes collected by the electric fields in the 
depleted region can drift to the p+ electrode and 
form a signal, 
a fraction of the holes will be trapped or recombined 
in the N-type region and cause an incomplete charge collection.

The number of produced charge carriers in 
semiconductor Ge is at $\mathcal{O}(10^5)$ for a 
$\mathcal{O}$(MeV) scale energy deposition. 
Although scattering and absorption 
of single holes are stochastic,
for a large number of holes,
their transportations in the N-type layer can be 
modeled by a transportation equation:

\begin{eqnarray}
\frac{\partial p}{\partial t}=D_{p}\bigtriangledown^2p-
\mu_p\bigtriangledown (\xi \cdot p) -\frac{p}{\tau_p}
\label{eq:1}
\end{eqnarray}

Where $p$ is the density of holes. 
The change rate of hole density ($\frac{\partial p}{\partial t}$) 
is determined by three processes: 
(i) the diffusion of holes ($D_{p}\bigtriangledown^2p$): 
$D_p$ the diffusion coefficient and 
$\bigtriangledown^2p$ the second spatial gradient of hole density, 
(ii) the drift of holes ($\mu_p\bigtriangledown (\xi \cdot p)$): 
$\mu_p$ the mobility of holes, 
$\bigtriangledown (\xi \cdot p)$ the spatial gradient of electric field 
$\xi$ and hole density p, 
(iii) the termination of holes ($\frac{p}{\tau_p}$), 
$\tau_p$ is the hole average lifetime.

\subsection{\label{sec:2.1} Li concentration profile}
During the fabrication of the Li-diffused layer, 
Li atoms are deposited on the surface of the Ge crystal by 
evaporation or electrolytic processes \cite{bib:36,bib:20,bib:21}. 
Then the crystal is heated in an annealing process
to facilitate the diffusion of Li atoms into the Ge lattice. 
The concentration profile of Li impurities ($C_{Li}$) 
depends on the annealing temperature ($T_{an}$), 
annealing time ($t_{an}$), 
and surface concentration 
($C_s$, usually saturated and equal to the 
solubility of Li in germanium).
The Li concentration profile is 
modeled by an erfc function \cite{bib:22}:

\begin{equation}
    C_{Li}(x)=C_s\cdot 
    erfc\left ( \frac{x}{2\sqrt{D_{Li}t_{an}}} \right )
\label{eq:2}
\end{equation}

\begin{equation}
    D_{Li}=D_0\cdot exp\left ( -\frac{H}{RT_{an}} \right )
\label{eq:3}
\end{equation}

Where x is the depth into the crystal surface. 
$D_{Li}$ is the diffusivity of Li in germanium 
in the unit of cm$^2/s$, 
R is the gas constant (R=1.98 cal/K). 
$D_0$ and H are diffusivity constant and activation energy, 
respectively. Their values are taken from measurements by 
Fuller: for $T_{an}$ in 473$\sim$873 K and 873$\sim$1273 K, 
$D_0$ as $2.5\times10^{-3}$ cm$^2$/s and $1.3\times10^{-3}$ cm$^2$/s, 
$H$ as 11800 cal and 10700 cal, respectively 
\cite{bib:22,bib:23}.

\begin{figure}[!htb]
\includegraphics
[width=1.0\hsize]
{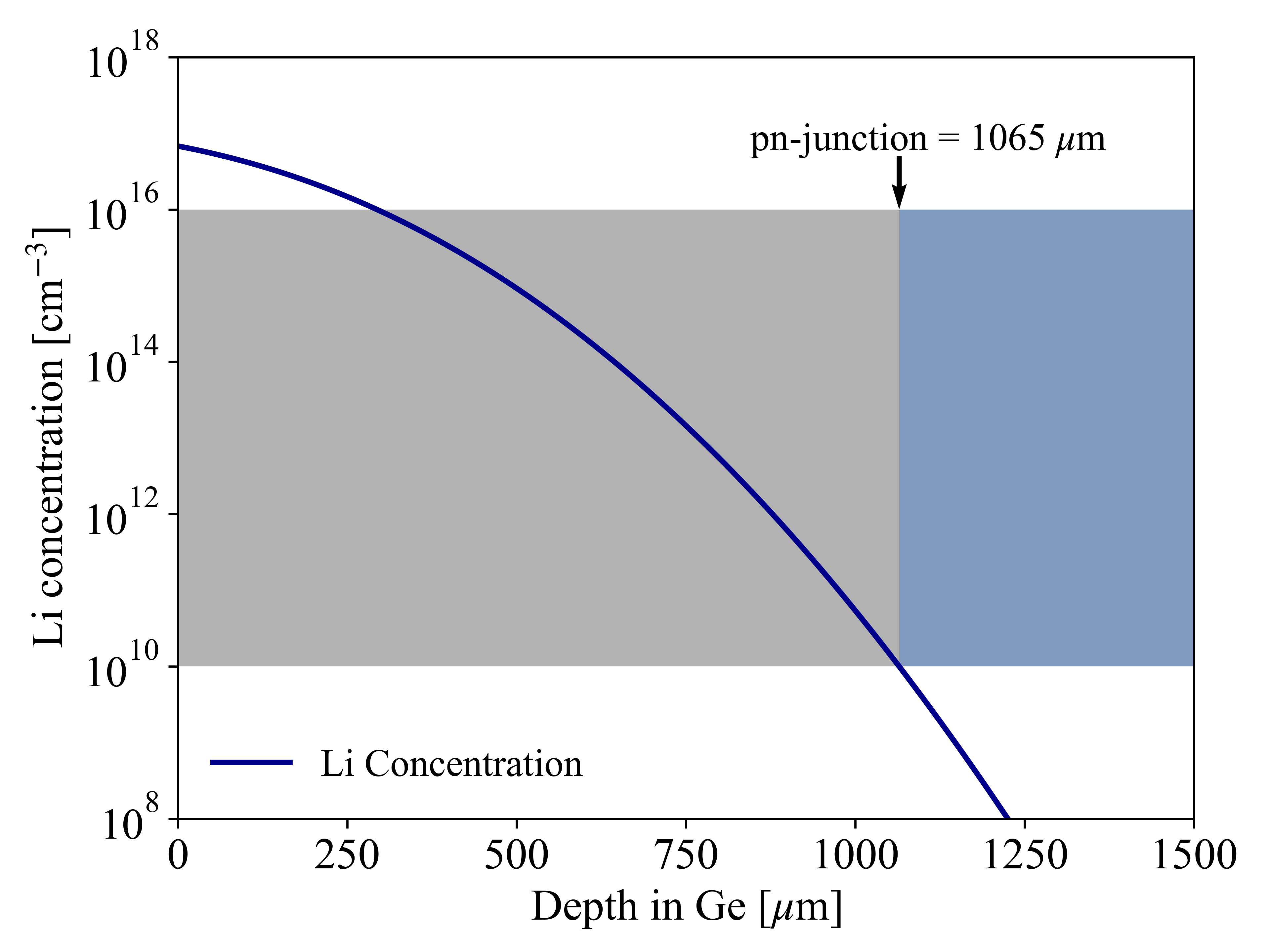}
\caption{\label{fig:LiProfile} Example of Li concentration profile 
in the surface layer, assuming a 573 K annealing temperature 
and 45-minute annealing time. 
The acceptor impurity is set to 10$^{10}$ cm$^{-3}$, 
the N-type and P-type regions are labeled in gray and 
blue respectively.}
\end{figure}

Figure.\ref{fig:LiProfile} demonstrates a typical
Li concentration profile in the crystal surface layer. 
Assuming the annealing is carried out in 573 K for 45 minutes, 
the saturated concentration of Li in Ge at this temperature 
is $C_s=6.8\times10^{16}$ cm$^{-3}$ \cite{bib:24}.
The Li concentration profile is calculated via Eq.\ref{eq:2}
and yields a 1065 $\mu$m thick N-type layer for a 
typical acceptor density of $10^{10}$ cm$^{-3}$. 

\subsection{\label{sec:2.2} Electric fields in the N-type layer}
When a reversed bias voltage is applied in the P-type HPGe detector, 
the P-type and a part of the N-type region near the 
detector surface are fully depleted. 
In the depleted region, the electric fields remove electrons 
and holes and leave the ionized impurities as space charges. 
The space charges and the applied bias voltage form 
electric fields in the N-type depleted region. 
The electric fields are calculated via solving 
the Poisson equation and taking the electric field at 
the pn-junction ($\xi_{pn}$) as the boundary condition:

\begin{equation}
    \xi_x=\xi_{pn}-\int_{x}^{x_{pn}}
    \frac{q}{\varepsilon}\left ( C_{Li}-C_a \right ) dx
\label{eq:4}
\end{equation}

\begin{figure}[!htb]
\includegraphics
[width=1.0\hsize]
{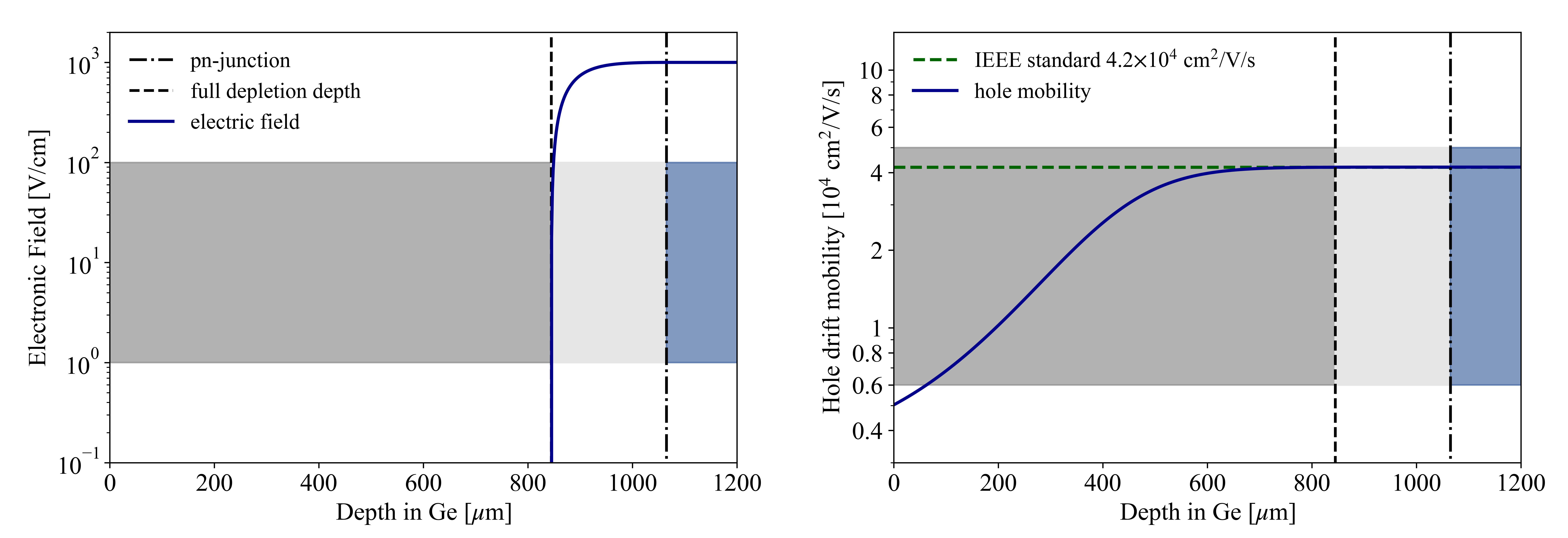}
\caption{\label{fig:EFieldandMobility} left: the electric field in 
the surface layer; right: the hole charge drift mobility 
$\mu_p$ in the surface layer, the crystal temperature 
is set at 90 K. Both figures use the Li profile in Fig.\ref{fig:LiProfile}. 
The neutral and depleted N-type regions are labeled in dark and light gray, 
the P-type regions are labeled in blue}
\end{figure}

Where $x_{pn}$ is the pn-junction depth in the crystal surface, 
$C_a$ is the concentration of the ionized acceptor impurity, 
q is the unit charge (q=1.6×10$^{-19}$ C), 
and $\varepsilon$ is the dielectric constant of Ge. 
The left panel of Fig.\ref{fig:EFieldandMobility} shows the electric fields 
in the N-type region using the Li profile in Fig.\ref{fig:LiProfile} 
and assuming a $\xi_{pn}$ of 1000 V/cm. 
The rapid decrease of the electric field near the 
full depletion depth (FDD) is mainly due to the 
steep profile of Li concentration and leaving 
the majority of the N-type region as a neutral layer 
without electric fields.

\subsection{\label{sec:2.3} Hole mobility and lifetime}
The hole charge drift mobility $\mu_p$ depends on the 
scattering of charge carrier holes by the impurity atoms and 
defects in Ge crystal. Three scattering processes are considered 
the main contributors to the hole charge drift mobility: 
ionized impurity scattering, neutral impurity scattering, 
and acoustic phonon scattering \cite{bib:26}. 
Other scattering processes, e.g., the optical phonon scattering 
and the defects scattering, are negligible compared to the 
three dominant processes for a P-type HPGe detector at a 
low-temperature working condition \cite{bib:26}, 
therefore are not considered in the calculation of $\mu_p$.

The ionized impurity scattering mobility $\mu_I$ is calculated 
by the BH model \cite{bib:27}, the neutral impurity scattering mobility 
$\mu_N$ by Sclar’s work \cite{bib:28}, and the acoustic phonon 
scattering mobility $\mu_A$ by the Bardeen and Shockley’s model \cite{bib:29}. 
All constants in the equations are replaced with their 
corresponding values, and the hole charge drift mobility can be 
expressed as:

\begin{equation}
    \frac{1}{\mu_p}=
    \frac{1}{\mu_I}+\frac{1}{\mu_N}+\frac{1}{\mu_A}
\label{eq:5}
\end{equation}

\begin{equation}
    \mu_I=
    \frac{(2.35\times10^{17}\cdot T^{3/2})/N_i}
    {ln\left ( (9.13\times10^{13}\cdot T^2)/N_i \right )} +
    \frac{(1.51\times10^{18}\cdot T^{3/2})/N_i}
    {ln\left ( (5.8\times10^{14}\cdot T^2)/N_i \right )}
\label{eq:6}
\end{equation}

\begin{equation}
    \mu_N=
    \frac{4.46\times10^{29}}{N_n}\cdot 
    \left ( T^{1/2}+4.28\cdot T^{-1/2} \right )
\label{eq:7}
\end{equation}

\begin{equation}
    \mu_A=7.77\times10^7\cdot T^{-3/2}
\label{eq:8}
\end{equation}

Where the $\mu_I$, $\mu_N$, $\mu_A$, and $\mu_p$ are in cm$^2$/V/s. 
$N_i$ and $N_n$ are the densities of ionized and 
neutral impurity (cm$^{-3}$), respectively. 
T is the temperature of Ge crystal (K). 
A flat profile of neutral impurity is assumed in this work, 
and $N_n$ is calculated by matching $\mu_p$ with the 
IEEE standard 4.2$\times$10$^4$ cm$^2$/V/s \cite{bib:30} 
in the P-type region. 
The right panel of Fig.\ref{fig:EFieldandMobility} 
demonstrates the hole charge drift mobility versus 
the depth into the surface, 
taking the Li concentration profile in Fig.\ref{fig:LiProfile}.
Near the surface of the detector, 
$\mu_p$ decreases significantly as the concentration of 
ionized impurity Li increases.

The diffusion coefficient $D_p$ can be calculated 
by $\mu_p$ via the Einstein relation: 
$D_p=(k_BT/q)\cdot\mu_p$. 
Where $k_B$ is the Boltzmann constant, 
and T is the crystal temperature. 
In the N-type surface layer, holes may be trapped by 
the impurities and vanish in various recombination processes 
before they are collected by the electric fields and 
transported to the P-type region \cite{bib:31}. 
Due to unclear types and concentration profiles of 
the trapping and recombination centers, 
a constant hole lifetime $\tau_p$ in the N-type layer and 
immediate recombination at the detector surface are 
assumed in this work. 
The effects of the electric field in the depleted region
is evaluated by setting $\tau_p$ proportional to electric field (maximum 
$\tau_p$ set to 100 $\mu$s), the charge collection efficiency (CCE) is similar to
that from a constant $\tau_p$. The differences in CCEs are within 0.6\%.
For a given detector, 
$\tau_p$ can be decided by comparing the calculated 
charge collection efficiency with measurements.

\subsection{\label{sec:2.4} Calculation of charge collection efficiency}
The charge collection efficiencies in different 
depths of the N-type layer can be calculated by 
solving Eq(\ref{eq:1}) via a numerical method. 
The parameters needed to solve Eq(\ref{eq:1}) are discussed in 
Sec\ref{sec:2.1}$\sim$\ref{sec:2.3} and are listed in 
Table \ref{tab:parModel}.

\begin{table}
    \centering
    \caption{\label{tab:parModel}Main parameters in the 
    Inactive Layer Model. 
    The Pre-Set parameters are related to the 
    production processes and crystal properties, 
    therefore are specified to a single detector. 
    The calculated parameters are determined by the 
    pre-set parameters using models 
    in Sec\ref{sec:2.1}$\sim$\ref{sec:2.3}}
    \renewcommand\arraystretch{1.5}
    \begin{tabular}{lll}
    \hline
        Parameter & Type & Note\\
    \hline    
        \multirow{2}{*}{Anneal temperature ($T_{an}$)}&
        \multirow{2}{*}{Pre-Set} &
        Detector specified,\\
        &&typically 473 K $\sim$ 673 K \\
        \multirow{2}{*}{Anneal time ($t_{an}$)} &
        \multirow{2}{*}{Pre-Set} &
        Detector specified,\\
        &&$\mathcal{O}(min)$$\sim\mathcal{O}(hour)$ \\
        \multirow{2}{*}{Li surface concentration ($C_s$)} &
        \multirow{2}{*}{Calculated} &
        \multirow{2}{*}{Solubility of Li in Ge at $T_{an}$} \\
        &&\\
        \multirow{2}{*}{Acceptor impurity} &
        \multirow{2}{*}{Pre-Set} &
        Detector specified \\
        &&$\mathcal{O}$(10$^{10}$ cm$^{-3}$)\\
        %
        \multirow{2}{*}{Crystal temperature ($T$)} &
        \multirow{2}{*}{Pre-Set} &
        Detector specified, \\
        &&typically 77 K $\sim$ 90 K \\
        Electric field &
        \multirow{2}{*}{Pre-Set} &
        Determined by the bias \\
        at the pn-junction ($\xi_{pn}$)&&
        voltage and geometry\\
        \multirow{2}{*}{Electric field ($\xi$)} &
        \multirow{2}{*}{Calculated} &
        By Li profile and electric field \\
        &&at pn-junction $\xi_{pn}$\\
        \multirow{2}{*}{Hole Mobility ($\mu_p$)} &
        \multirow{2}{*}{Calculated} &
        \multirow{2}{*}{By Li profile} \\
        &&\\
        \multirow{2}{*}{Diffusion Coefficient ($D_p$)} &
        \multirow{2}{*}{Calculated} &
        \multirow{2}{*}{By mobility $\mu_p$} \\
        &&\\
        \multirow{2}{*}{Hole lifetime ($\tau_p$)} &
        \multirow{2}{*}{Pre-Set} &
        Detector specified, \\
        &&$\mathcal{O}(ns)\sim\mathcal{O}(\mu s)$ \\
        \hline
    \end{tabular}
\end{table}

As the collection of holes mainly depends on the 
maximum depth holes can reach before they are trapped 
or recombined, the Eq.\ref{eq:1} is processed into a 1-D
differential equation:

\begin{eqnarray}
    \label{eq:9}
    p(i,t+1)&&=C_{diff}(i+1)\cdot p(i+1,t)\\ \nonumber
    &&+\left [C_{diff}(i-1)+C_{drif}(i-1)\right ]\cdot p(i-1,t) \\ \nonumber
    &&+\left [1-2C_{diff}(i)-C_{drif}(i)-C_{Te}(i)\right ]\cdot p(i,t)
\end{eqnarray}

The $p(i,t)$ is the hole density of $i^{th}$ bin at time $t$, 
$C_{diff}(i)=D_p(i)/\triangle x^2\cdot \triangle t$ 
describes the diffusion of holes, 
$C_{drift}(i)=\mu_p(i)\cdot \xi_x(i)/\triangle x\cdot \triangle t$
describes the drift of holes, and
$C_{Te}(i)=\triangle t/\tau_p(i)$
is the fraction of holes been trapped or recombined 
in the ith bin during $\triangle t$ times. 
The Eq.\ref{eq:9} is calculated with a 
space bin-size $\triangle x$ of 20 $\mu$m and a 
time bin-size $\triangle t$ of 0.1 ns.

The solving of Eq.\ref{eq:9} starts with the injection of 
holes at depth $x$, $p(i=x,t=0)=N_0$ and 
the rest of bins are set as empty. 
During the transportation, holes reaching the deepest 
bin (located at the pn-junction) are removed and counted 
as collected. The CCE of depth $x$ is calculated as the 
ratio of holes collected during the signal 
forming time of the detector.

\section{Result and discussion}
\subsection{Charge collection efficiency of the CDEX-1B detector}

The inactive layer model is applied to the CDEX-1B detector
to calculate the CCE function.
The model parameters specified for 
the CDEX-1B detector are listed in Table\ref{tab:C1B}. 
The annealing process can be carried out in wide ranges 
of temperature (473 $\sim$ 673 K) and 
time (a few minutes $\sim$ several hours) 
depending on the process \cite{bib:20,bib:21,bib:32,bib:33}. 
We take a typical 573 K annealing temperature and 
45 minutes of annealing time for the CDEX-1B detector. 
The $C_s$ are taken as the saturated concentration of Li 
in Ge at 573 K \cite{bib:24}. 
The acceptor impurity in the detector is evaluated in our 
previous work \cite{bib:34}
by matching the simulated pulse with measurement.
An average acceptor impurity of 
1.39$\times$10$^{10}$ cm$^{-3}$ in the N-type layer is adopted.
According to an electric field simulation \cite{bib:34,bib:35}, 
the electric fields ($\xi_{pn}$) are in the range of 
500 to 1500 V/cm for the lateral surface of the CDEX-1B detector. 
The hole lifetime ($\tau_p$) in the inactive layer is in 
$\mathcal{O}$(ns)$\sim\mathcal{O}$($\mu$s) scale
for a typical P-type HPGe detectors\cite{bib:20,bib:21}.
We run a scan of ($\tau_p$) in 1 ns$\sim$5 $\mu$s range,
and the modeled CCE functions of different ($\tau_p$) 
are compared with the empirical function.
The best match result corresponds 
to a hole lifetime of 0.40 $\mu$s. 

\begin{table}
    \centering
    \caption{\label{tab:C1B}Inactive layer parameters of two P-type HPGe detectors}
    \begin{tabular}{ccc}
    \hline
        Parameter & CDEX-1B & PSPGe\\
    \hline    
        Anneal temperature ($T_{an}$) & 573 K & 623 K \\
        Anneal time ($t_{an}$) & 45 min & 18 min \\
        Li surface concentration (C$_s$) &
        6.8$\times$10$^{16}$ cm$^{-3}$ & 1.2$\times$10$^{17}$ cm$^{-3}$\\
        Acceptor impurity& 
        1.39$\times$10$^{10}$ cm$^{-3}$ & 0.30$\times$10$^{10}$ cm$^{-3}$\\
        Crystal temperature (T) & 90 K & 90 K\\
        Electric field at pn-junction ($\xi_{pn}$)&
        1000$\pm$500 V/cm & 500$\pm$300 V/cm\\
        Hole lifetime ($\tau_p$) & 0.40 $\mu$s & 0.41 $\mu$s\\
        \hline
    \end{tabular}
\end{table}

\begin{figure}[htb]
    \includegraphics
    [width=1.0\hsize]
    {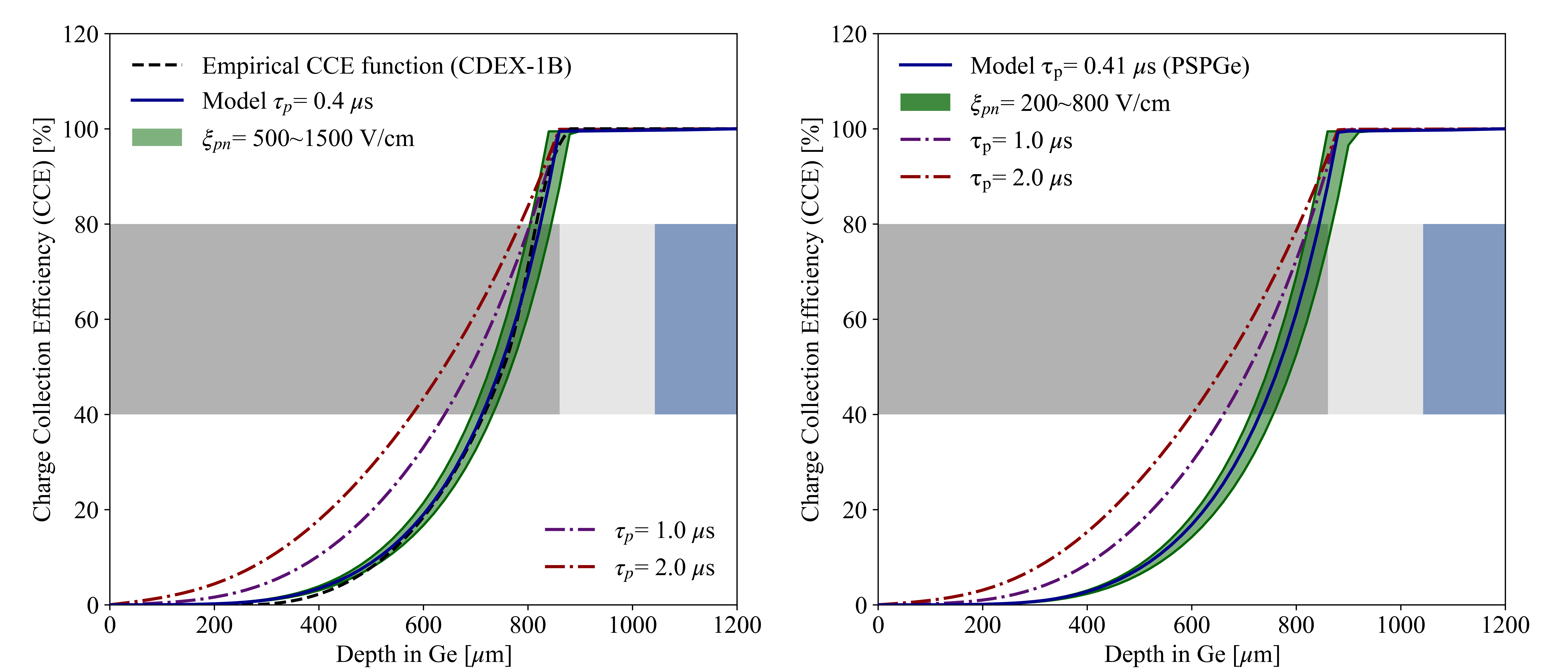}
    \caption{\label{fig:CCE-C1B} Charge collection 
    efficiencies for CDEX-1B detector (Left) and PSPGe detector (Right). 
    CCE of CDEX-1B is compared with the empirical CCE function 
    (black dash line in left figure). The green region indicates 
    the variations of CCEs with different $\xi_{pn}$, 
    the red and purple dash lines are CCEs with 
    different hole lifetime $\tau_p$. 
    The N-type neutral region, 
    N-type depleted region and P-type region are 
    labeled in dark gray, light gray and blue, respectively.
    }
\end{figure}

The calculated CCE function is shown with the 
empirical function in Fig.\ref{fig:CCE-C1B} 
The best-matched result with $\xi_{pn}$=1000 V/cm 
yields a full charge collection depth (99\% CCE) 
of 859 $\mu$m. 
The uniformity of the inactive layer thickness is evaluated
by the variations of electric field $\xi_{pn}$.
For $\xi_{pn}$ in 500$\sim$1500 V/cm,
the corresponding FCCDs are in the range of 
839$\sim$884 $\mu$m and are consistent with the result from 
the previous measurement (850$\pm$120 $\mu$m). 
The CCEs calculated by the inactive model and the 
empirical function are also in good agreement.

\subsection{Model validation in a P-type semi-planar HPGe detector}

\begin{figure}[!htb]
    \includegraphics
    [width=1.0\hsize]
    {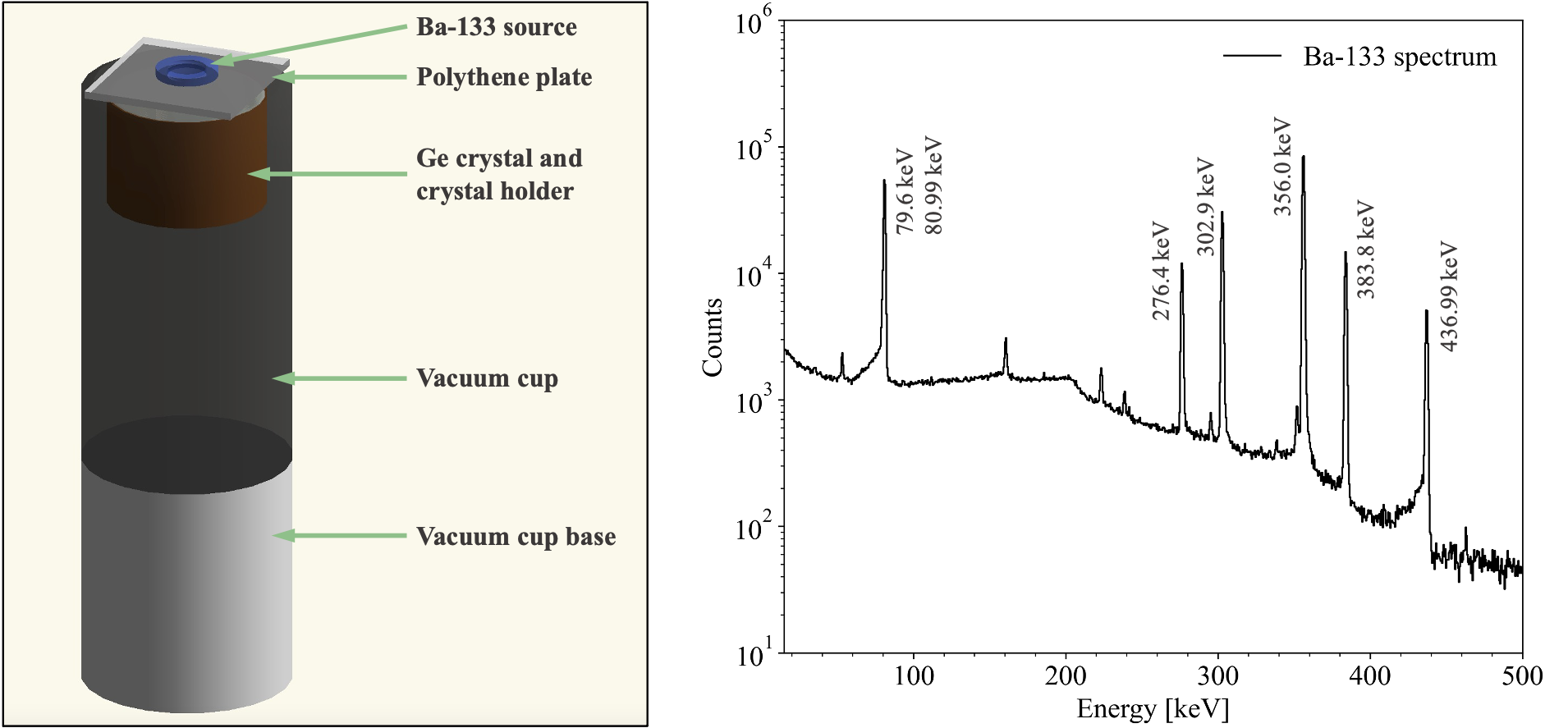}
    \caption{\label{fig:spBa133} 
        Set up of the Ba-133 source measurement (Left)
        and the measured Ba-133 spectrum (Right)
    }
\end{figure}

The inactive layer model is applied to a 
P-type semi-planar Ge (PSPGe) detector purchased from ORTEC.
The detector crystal has a diameter of 80 mm, a height of 42.6 mm 
and a small size p+ contact. 
The crystal surface is formed by a Li-diffused N-type layer.
The thickness of the surface inactive layer is measured by a Ba-133 source
following the method described in \cite{bib:16}.
Fig.\ref{fig:spBa133} shows the setup of the measurement and the measured spectrum of the Ba-133 source.
The measured FCCD is 870$\pm$67 $\mu$m.

\begin{figure}[!htb]
    \includegraphics
    [width=1.0\hsize]
    {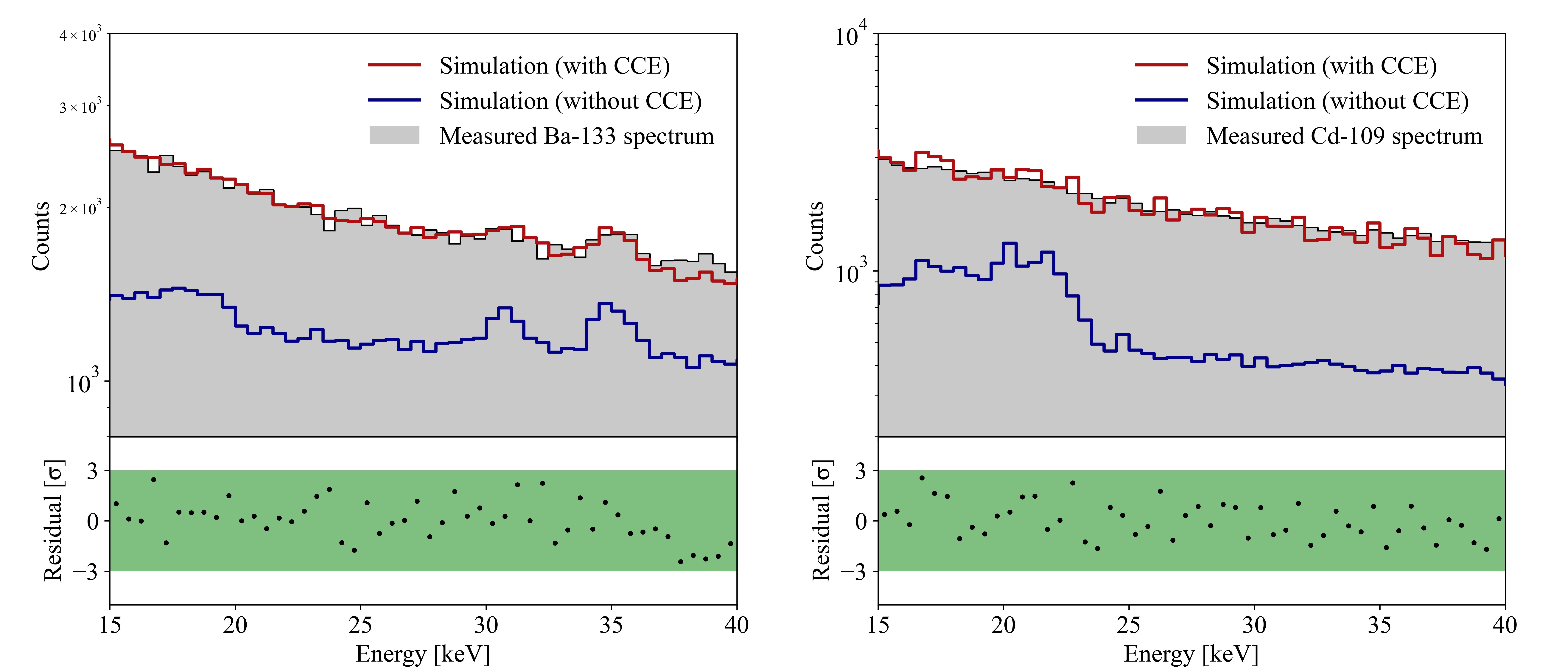}
    \caption{\label{fig:valBa133Cd109} 
        Simulated and measured spectra in 15$\sim$40 keV region
        for Ba-133 (Left) and Cd-109 (Right) sources.
        Simulated spectra without CCE correction are labeled in blue 
        (energy deposited in the inactive layer not recorded), spectra after
        CCE correction are labeled in red. Normalized residuals are shown below the spectra.
    }
\end{figure}

The charge collection efficiency (CCE) is calculated for the PSPGe detector
using the parameters listed in Table.\ref{tab:C1B} and shown in Fig.\ref{fig:CCE-C1B}. 
Simulated spectra are generated using GEANT4\cite{bib:37} and the 
calculated CCE: energy depositions in the inactive layer are corrected by the CCE
according to their positions.
The hole lifetime ($\tau_p$) is estimated by comparing the 
simulated spectra with the measured one,
and the best matched $\tau_p$ is 0.41 $\mu$s for $\xi_{pn}$=500 V/cm.
The calculated FCCDs are 879 $\mu$m for $\xi_{pn}$=500 V/cm and 859$\sim$918 $\mu$m 
for $\xi_{pn}$ in range of 200$\sim$800 V/cm.
As shown in Fig.\ref{fig:valBa133Cd109},
simulated spectra of Ba-133 and Cd-109 sources
after CCE correction are in good agreement with measurements in 15$\sim$30 keV region
where the spectrum shape is significantly affected by the incomplete charge collection in the inactive layer.

\subsection{Effects of Li doping processes on the charge collection efficiency}
\begin{figure}[!htb]
    \includegraphics
    [width=1.0\hsize]
    {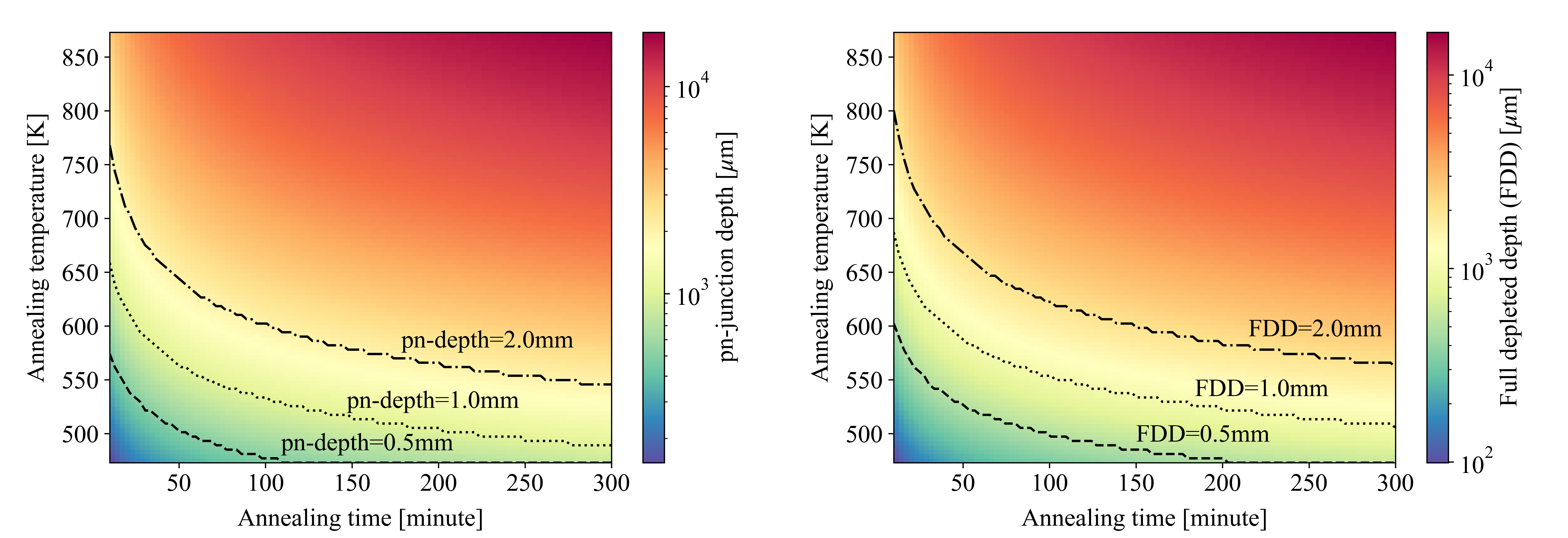}
    \caption{\label{fig:pnFDD} pn-junction depths (Left) 
    and full depleted depths (Right) under different 
    annealing temperatures and times. 
    The full depleted depths are calculated assuming 
    $\xi_{pn}$ = 1000 V/cm.}
\end{figure}

The annealing time and temperature varies for different 
Li doping processes \cite{bib:20,bib:21}. 
To assess the influence of different annealing parameters
on the structure and charge collection efficiency of the 
inactive layer, we calculate
the depths of the pn-junction (pn-depth) and the 
full depleted region (FDD) under different 
annealing temperatures and times, 
as shown in Fig.\ref{fig:pnFDD}.

\begin{figure}[!htb]
    \includegraphics
    [width=1.0\hsize]
    {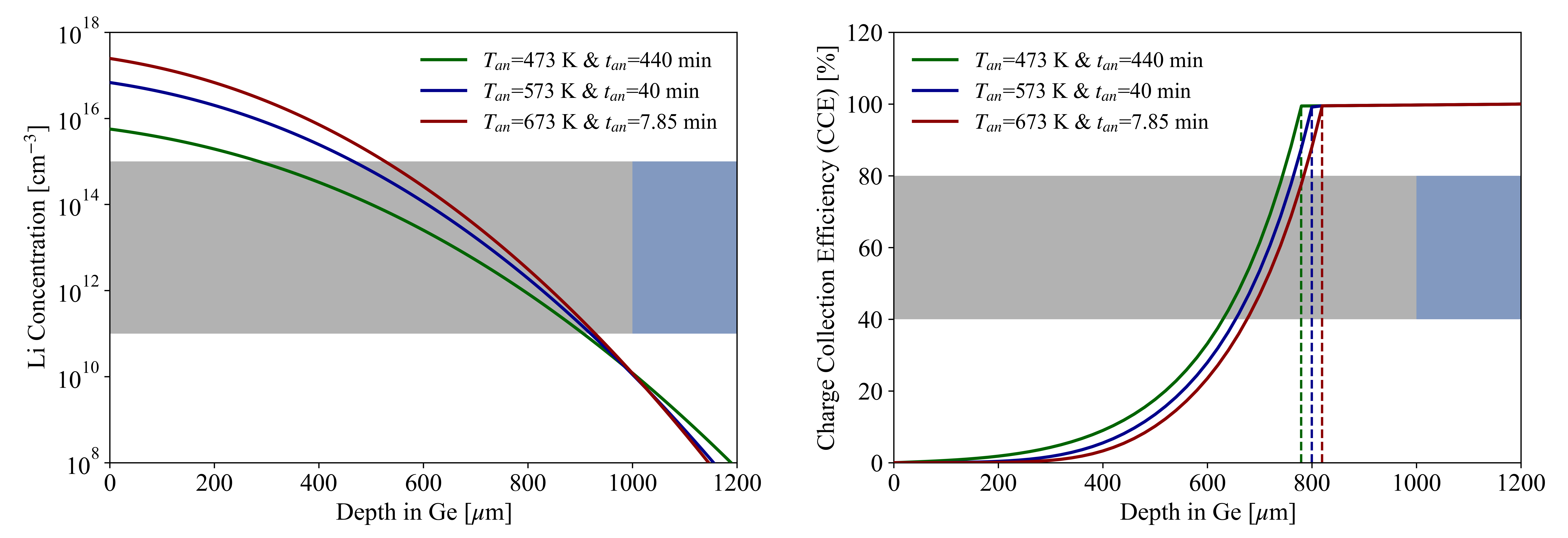}
    \caption{\label{fig:pnCompare} Li concentration profiles 
    (left) and charge collection efficiencies (right) of 
    different annealing conditions with identical 1 mm 
    pn-junction depths. The positions of the full 
    charge collection depths (99\% CCE) are marked 
    by the dash lines for all three annealing conditions.}
\end{figure}

\begin{figure}[!htb]
    \includegraphics
    [width=1.0\hsize]
    {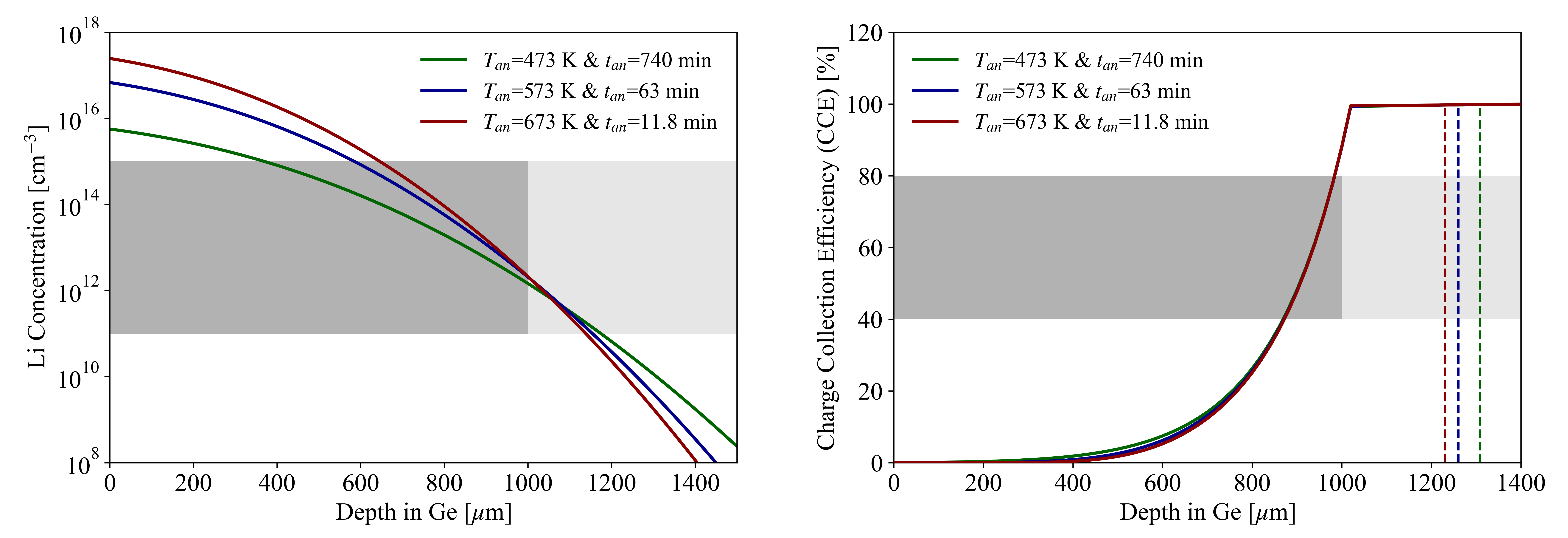}
    \caption{\label{fig:FDDCompare} Li concentration profiles 
    (left) and charge collection efficiencies (right) 
    of different annealing conditions with identical 1 mm 
    full depleted depths. The positions of the pn-junctions 
    are marked by the dash lines for all three annealing 
    conditions.}
\end{figure}

Different annealing processes could produce identical 
pn-junction depths or full depleted depths. 
However, the charge collection efficiencies and the 
full charge collections depth may differ due to the 
difference in Li concentration profiles in the N-type layer. 
Annealing processes with identical pn-junction depths (1 mm) 
and FDD (1 mm) are selected, and the corresponding CCEs are 
calculated assuming an electric field $\xi_{pn}$ =1000 V/cm 
and a hole lifetime $\tau_p$ = 0.4 $\mu$s. 
The results are shown in Fig.\ref{fig:pnCompare} 
and Fig.\ref{fig:FDDCompare}.

As shown in Fig.\ref{fig:pnCompare}, 
the thickness of the inactive layer 
(define as the full charge collection depth) and 
charge collection efficiencies vary from different 
annealing temperatures under the same pn-junction depth. 
A higher annealing temperature corresponds to lower 
charge collection efficiencies and a thicker inactive layer 
due to a thinner full depleted region and lower charge 
carrier mobilities in the N-type region.

As shown in Fig.\ref{fig:FDDCompare}, 
different annealing processes with similar 
full depleted depths have presented approximative 
full charge collection depths and CCE profiles. 
Due to the steep profile of the electric fields 
(as shown in Fig.\ref{fig:EFieldandMobility}) 
facilitating the collections of holes in the 
full depleted region, the CCE quickly reaches $\sim$100\% 
in the full depleted region. 
It shows that the full depleted depth is a better 
parameter to describe the thickness of the inactive 
layer compared to the depth of the pn-junction.

\section{Conclusions}
A 1-D model describes the charge collections in the Li-diffused 
inactive layer of the P-type HPGe detector
has been developed in this work. The charge collection efficiencies in 
different depths of the inactive layer can be calculated 
by solving the hole transportation equation. 
The effects of Li doping process, the annealing temperature, 
and the annealing time on the charge collection efficiency 
are evaluated. And the assessment shows
the structure of inactive layer varies according to 
the Li-diffusion process and
the thickness of the inactive layer mainly 
depends on the full depleted depth rather than the position 
of the pn-junction.

The charge collection efficiencies of the CDEX-1B (PPCGe) detector is
calculated using the inactive layer model. 
The model is validated in another P-type semi-planar Ge detector.
The full charge collection depth and charge collection 
efficiencies in different depths for both detectors agree with measurements.

The implementation of this model requires some detector 
specified parameters, e.g., the annealing time and 
temperature of the Li-diffusion process, 
recording those parameters during the fabrication can 
help to build a more precise model of the inactive layer. 
The hole lifetime in the inactive layer is a fairly 
complicated matter, we take a simple constant assumption 
in this work for lack of detail knowledge on the 
recombination and trapping processes. 
An average hole lifetime ($\tau_p$) in the inactive layer
can be estimated by
scanning $\tau_p$ to match the simulated spectra with measurements.

The model in this work can be easily extended 
to other types of P-type
Ge detector (ICPC, BEGe, Coaxial$\dots$) 
by adjusting the detector specified parameters in the model.

\section*{Acknowledgments:}
This work was supported by the 
National Key Research and Development Program of China 
(Grant No. 2017YFA0402200) 
and the National Natural Science Foundation of China 
(Grant No. 12175112, No. 12005111, and No. 11725522).






\begin{thebibliography}{99}
%

\bibitem{bib:1}
R. D. Baertsch and N. R. Hall.
IEEE Trans Nucl Sci. {\bf17}, 3, 235-240, (1970).

\bibitem{bib:2}
W. L. Hansen.
Nucl. Instrum. Methods. {\bf94}, 377-380 (1971).

\bibitem{bib:3}
E.E. Haller.
Mater Sci Semicond Process. {\bf9}, 4, 408-422 (2006).

\bibitem{bib:4}
L. T. Yang, \emph{et al}.
Chinese Phys. C. {\bf42}, 023002 (2018).

\bibitem{bib:5}
H. Jiang, \emph{et al}.
Phys. Rev. Lett. {\bf120}, 241301 (2018).

\bibitem{bib:6}
A. K. soma, \emph{et al}.
Nucl. Instrum. Methods. Phys. Res. A. {\bf836}, 67 (2016).

\bibitem{bib:7}
N. Abgrall, \emph{et al}.
(Majorana Collaboration),
Phys. Rev. Lett. {\bf118}, 161801 (2017).

\bibitem{bib:8}
M. Agostini, \emph{et al}.
(GERDA Collaboration),
Phys. Rev. Lett. {\bf125}, 011801 (2020).

\bibitem{bib:9}
M. Agostini, \emph{et al}.
(GERDA Collaboration),
Phys. Rev. Lett. {\bf125}, 252502 (2020).

\bibitem{bib:10}
S. I. Alvis, \emph{et al}.
(Majorana Collaboration),
Phys. Rev. C {\bf100}, 025501 (2019).

\bibitem{bib:11}
E. Aguayo, \emph{et al}.
Nucl. Instrum. Methods A. {\bf701}, 176–185 (2013).

\bibitem{bib:12}
H. Jiang, \emph{et al}.
Chinese Phys. C. {\bf40}, 096001 (2016).

\bibitem{bib:13}
L.T. Yang, \emph{et al}.
Nucl. Instrum. Methods A. {\bf886}, 13-23 (2018).

\bibitem{bib:14}
H.B. Li, \emph{et al}.
Astropart. Phys. {\bf56}, 1-8 (2014).

\bibitem{bib:15}
B. Lehnert.
Journal of Physics: Conference Series {\bf718}, 062035 (2016).

\bibitem{bib:16}
J.L. Ma, \emph{et al}.
Appl Radiat Isot. {\bf127}, 130-136 (2017).

\bibitem{bib:17}
W. Shockley.
J. Appl. Phys. {\bf9}, 635 (1938).

\bibitem{bib:18}
S. Ramo.
Proceedings of the I.R.E. {\bf584}, (1939).

\bibitem{bib:19}
R.J. Cooper, \emph{et al}.
Nucl. Instrum. Methods A, {\bf629}, 1 (2011).

\bibitem{bib:36}
B. Pratt and F. Friedman. 
J. Appl. Phys. {\bf37}, 1893 (1966).

\bibitem{bib:20}
Q. Looker. PhD Thesis,University of California, (2014).

\bibitem{bib:21}
B. Lehnert. PhD Thesis, Dresden University of Technology, (2016).

\bibitem{bib:22}
C. S. Fuller and J. A. Ditzenberger.
Phys. Rev. {\bf91}, 193 (1953).

\bibitem{bib:23}
C. S. Fuller and J. C. Severiens.
Phys. Rev. {\bf96}, 21 (1954).

\bibitem{bib:24}
G. H. R. Kegel, \emph{et al}.
J. Electrochem. Soc, {\bf118}, 10, 1662-1665 (1971).


\bibitem{bib:26}
H. Mei, \emph{et al}.
JINST, {\bf11}, P12021 (2016).

\bibitem{bib:27}
D. Chattopadhyay and H. J. Queisser.
Rev. Mod. Phys. {\bf53}, 745 (1981).

\bibitem{bib:28}
N. Sclar.
Phys. Rev. {\bf104}, 1559 (1956).

\bibitem{bib:29}
J. Bardeen and W. Shockley.
Phys. Rev. {\bf80}, 72 (1950).

\bibitem{bib:30}
S. Wagner, \emph{et al}.
IEEE, 1160–1993 (1993).

\bibitem{bib:31}
J. Lauwaert and P. Clauws.
Thin Solid Films, {\bf518}, 9, 2330-2333 (2010).

\bibitem{bib:32}
W. L. Hansen and E. E. Haller.
IEEE Trans Nucl Sci, {\bf28}, 1, 541-543 (1981).

\bibitem{bib:33}
R. Pehl, \emph{et al}.
IEEE Trans Nucl Sci, {\bf19}, 265-269 (1972).

\bibitem{bib:34}
R.M.J. Li, \emph{et al}.
NUCL SCI TECH {\bf33}, 57 (2022). 

\bibitem{bib:35}
Z. She, \emph{et al}.
J. Instrum. {\bf16}, T09005 (2021).

\bibitem{bib:37}
S. Agostinelli, \emph{et al}.
Nucl. Instrum. Methods A, {\bf3}, 506, 250-303 (2003).

\end{thebibliography}

\end{document}